\documentclass{article}
\PassOptionsToPackage{numbers, compress}{natbib}
 \usepackage[preprint]{neurips_2025}

\usepackage[utf8]{inputenc} % allow utf-8 input
\usepackage[T1]{fontenc}    % use 8-bit T1 fonts
\usepackage{hyperref}       % hyperlinks
\usepackage{url}            % simple URL typesetting
\usepackage{booktabs}       % professional-quality tables
\usepackage{amsfonts}       % blackboard math symbols
\usepackage{nicefrac}       % compact symbols for 1/2, etc.
\usepackage{microtype}      % microtypography
\usepackage{xcolor}         % colors

\usepackage{amsmath}
\usepackage{multirow}
\usepackage{float}
\usepackage{makecell}
\usepackage{graphicx}
\usepackage[misc]{ifsym}
\usepackage{enumitem}
\usepackage{wrapfig}
\usepackage{caption}

\definecolor{Blue}{rgb}{0.2,0.2,0.6}
\hypersetup{
  colorlinks=true,
  citecolor=Blue,
  linkcolor=Blue,
  urlcolor=Blue,
}

\title{FreeAudio: Training-Free Timing Planning for Controllable Long-Form Text-to-Audio Generation}

\author{
Yuxuan Jiang$^{1,2,*}$, 
Zehua Chen$^{1,2,*}$, 
Zeqian Ju$^{3,*}$, 
Chang Li$^{2,3}$, 
Weibei Dou$^1$, 
Jun Zhu$^{1, \textrm{\Letter}}$\\
$^1$Tsinghua University \; $^2$Shengshu AI \; $^3$University of Science and Technology of China \\
\textrm{\Letter} \texttt{dcszj@mail.tsinghua.edu.cn}
}
\begin{document}

\maketitle

\begin{abstract}
Text-to-audio (T2A) generation has achieved promising results with the recent advances in generative models. However, because of the limited quality and quantity of temporally-aligned audio-text pairs, existing T2A methods struggle to handle the complex text prompts that contain precise timing control, \textit{e.g.}, \textit{owl hooted at 2.4s-5.2s}. Recent works have explored data augmentation techniques or introduced timing conditions as model inputs to enable timing-conditioned 10-second T2A generation, while their synthesis quality is still limited. In this work, we propose a novel training-free timing-controlled T2A framework, FreeAudio, making the first attempt to enable \textit{timing-controlled long-form T2A generation}, \textit{e.g.}, \textit{owl hooted at 2.4s-5.2s and crickets chirping at 0s-24s}. Specifically, we first employ an LLM to plan non-overlapping time windows and recaption each with a refined natural language description, based on the input text and timing prompts. Then we introduce: 1) Decoupling \& Aggregating Attention Control for precise timing control; 2) Contextual Latent Composition for local smoothness and Reference Guidance for global consistency. Extensive experiments show that: 1) FreeAudio achieves state-of-the-art timing-conditioned T2A synthesis quality among training-free methods and is comparable to leading training-based methods; 2) FreeAudio demonstrates comparable long-form generation quality with training-based Stable Audio and paves the way for timing-controlled long-form T2A synthesis.
Demo samples are available at:~\url{https://freeaudio.github.io/FreeAudio/}.
\end{abstract}

\section{Introduction}

Audio generation is playing an increasingly important role in content creation. It not only generates vivid and realistic sound elements for environments to enhance the auditory immersion of narrative scenes, but also serves as a key component in the design of music and sound effects in games. Meanwhile, as an essential part of multimodal generative systems, audio content that is coherent and finely controllable is being widely adopted in multimedia productions, significantly enhancing their expressiveness and dissemination potential. This trend continues to expand the role of audio throughout the creative workflow. In recent years, Text-to-Audio (T2A) generation has been intensively explored to synthesize semantically aligned and acoustically realistic audio samples from natural language prompts~\cite{liu2023audioldm,ghosal2023text,huang2023make,yang2023uniaudio}. Benefiting from the advancement of generative models, T2A systems have demonstrated remarkable generation quality by capturing the distribution of collected large-scale audio samples.

Despite the impressive progress made by existing T2A models, they still face significant challenges in achieving precise timing control and long-term continuity. These models often struggle to accurately follow lengthy and complex prompts, which limits their practical performance and scalability~\cite{bovzic2024survey}. Extending the capabilities of current T2A models remains a key research direction in the field of audio generation. However, most publicly available audio datasets are relatively short (typically around 10 seconds) and only offer coarse-grained, subtitle-level annotations~\cite{kim2019audiocaps,gemmeke2017audio,chen2020vggsound,fonseca2021fsd50k}. These data limitations hinder the development of models capable of fine-grained timing control and long-form coherence. To address these challenges, prior work has explored various dimensions such as long-duration generation, temporal control, and event sequence modeling. For example, Stable Audio~\cite{evans2024fast} leverages diffusion models trained on large-scale proprietary long-form audio datasets to generate up to 4 minutes and 30 seconds of audio. However, its approach relies on high-cost, precisely labeled data and extensive computational resources, making it difficult to replicate in open environments. Make-an-Audio 2~\cite{huang2023make} enhances a model’s temporal sequencing ability through data augmentation strategies, while methods such as PicoAudio~\cite{xie2024picoaudio} and MC-Diffusion~\cite{guo2024audio} incorporate event timestamps as additional conditioning signals to help the model interpret and control temporal information. Nonetheless, these approaches often depend on pre-defined event categories and struggle to generalize to open-ended natural language descriptions, which limits their practicality and general applicability.

A straightforward approach to improving temporal control in T2A models is to construct large-scale, long-form datasets with precise temporal annotations and train a dedicated T2A model (e.g., Stable Audio~\cite{evans2024fast,evans2024long,evans2025stable}) that can better understand and execute complex instructions. However, such methods suffer from significant limitations: collecting and annotating high-quality, long-duration, and controllable datasets is prohibitively costly, and training or fine-tuning diffusion models at this scale demands substantial computational resources and engineering overhead~\cite{mei2024wavcaps,xie2025audiotime,xu2024towards}. To address these challenges, we propose FreeAudio, a text-to-audio generation framework that supports temporal control and long-form synthesis without additional training. Our goal is to enable a structured understanding of complex prompts (e.g., a combination of text and timing prompts) in order to facilitate controllable, high-quality audio synthesis in a training-free manner. Leveraging audio timestamps and corresponding sub-prompts, we introduce Decoupling \& Aggregating Attention Control to improve both the flexibility and quality of T2A generation. Specifically, we first employ the planning capability of large language models (LLMs)~\cite{hurst2024gpt,guo2025deepseek,gemini25pro} to convert textual and temporal prompts into a sequence of non-overlapping time windows, each accompanied by a natural language recaption~\cite{yang2024mastering}. Audio segments are then independently generated for each sub-prompt within its designated time window, and subsequently merged via truncation and alignment of effective audio segments. This Decoupling \& Aggregating Attention Control enables the generation of temporally aligned audio based on natural language descriptions. Furthermore, to extend temporal control to longer durations, we synchronize multiple diffusion processes to optimize a joint generation procedure that produces coherent and stylistically consistent outputs~\cite{dai2025latent,bar2023multidiffusion}. Concretely, we enhance local boundary smoothness across adjacent audio segments by explicitly incorporating contextual information in the latent space, while maintaining global stylistic consistency via reference-guided generation.

In summary, FreeAudio enables unified timing control and long-form audio generation. Compared with prior approaches, our main contributions are as follows:

\begin{itemize}[leftmargin=*]
\item We propose FreeAudio, the first training-free text-to-audio generation framework that simultaneously achieves precise timing control and consistent long-form synthesis. Without requiring any model training or parameter updates, our method can generate high-quality, temporally aligned, and structurally coherent audio content from complex natural language prompts.

\item We achieve fine-grained timing control and long-form generation through a two-stage approach. FreeAudio first leverages LLM planning and recaptioning to structure the generation process, followed by Decoupling \& Aggregating Attention Control to segment and coordinate generation for precise temporal alignment. To further ensure long-form consistency, we apply Contextual Latent Composition and Reference Guidance to enhance boundary smoothness and global coherence. Finally, decoded segments are trimmed and concatenated to produce long-form audio outputs.

\item We conduct comprehensive objective and subjective evaluations. On the AudioCondition test set, FreeAudio delivers comparable or superior performance in timing alignment and audio generation quality. On the AudioCaps and MusicCaps test sets, FreeAudio achieves better quality and consistency in 26-second and 90-second generation tasks compared to 10-second T2A models extended to longer durations. Results show that FreeAudio, without additional training data or supervision, achieves both precise temporal alignment and coherent long-duration synthesis.

\end{itemize}

\section{Related Work}

\subsection{Diffusion-Based Text-to-Audio Generation}

Diffusion models have emerged as a dominant approach for T2A generation, offering superior fidelity and flexibility. Diffsound~\cite{yang2023diffsound} first introduced discrete diffusion on mel-spectrogram tokens. Building on this, AudioLDM~\cite{liu2023audioldm} proposed latent diffusion conditioned on CLAP~\cite{wu2023large} embeddings, enabling higher audio quality while reducing dependence on paired data. Stable Audio advanced long-form audio generation by leveraging large-scale proprietary datasets to train diffusion models capable of producing high-quality audio over extended durations. While it introduced timing-conditioned embeddings, they are primarily used for controlling the overall length of the generated audio rather than providing fine-grained temporal guidance. On the semantic alignment front, TANGO~\cite{ghosal2023text} and TangoFlux~\cite{hung2024tangoflux} explored instruction-tuned language models and preference-based fine-tuning to improve alignment with complex text prompts and enhance compositional accuracy. Despite these advancements, existing diffusion-based models still face challenges in explicitly modeling long-range temporal structures and achieving fine-grained timing control within audio generation.

\subsection{Timing-Controlled Audio Generation}

Recent works have explored improving timing controllability in text-to-audio generation. Audio-Agent~\cite{wang2024audio} leverages large language models (LLMs) to decompose text or video inputs into timestamped instructions to guide pretrained diffusion models. PicoAudio~\cite{xie2024picoaudio} integrates temporal information directly into model design, supported by fine-grained, temporally aligned audio-text datasets. MC-Diffusion~\cite{guo2024audio} enhances controllability by introducing timestamped content and style conditions via a trainable control encoder. TG-Diff~\cite{du2024controllable} proposes a training-free method that adjusts latent variables during inference to enforce temporal alignment. AudioComposer~\cite{wang2025audiocomposer} achieves fine-grained control of audio generation purely from natural language descriptions using cross-attention-based diffusion transformers. Despite these advances, scalable and precise timing control in open-domain scenarios and fine-grained control in long-form audio generation remain open challenges.

\subsection{Long-Form Audio Generation}

Long-form audio generation methods can be broadly categorized into autoregressive modeling and diffusion-based modeling. AudioGen~\cite{kreuk2022audiogen} and MusicGen~\cite{copet2023simple} adopt transformer-based architectures to model discrete representations of audio and music, enabling coherent generation conditioned on text or melody. AudioLM~\cite{borsos2023audiolm} introduces a hybrid tokenization framework to jointly capture local fidelity and long-term structure, achieving high-quality speech and music continuations. MusicLM~\cite{agostinelli2023musiclm} further extends autoregressive modeling with a hierarchical design, supporting controllable music generation over extended durations. In contrast, diffusion-based approaches model audio generation as iterative denoising over latent spaces. AudioLDM 2~\cite{liu2024audioldm} integrates a self-supervised AudioMAE encoder~\cite{huang2022masked}, a GPT-2-based language model, and a latent diffusion model, extending the capability of audio generation to longer durations. Stable Audio~\cite{evans2024fast} incorporates timing embeddings into latent diffusion models, enabling controllable synthesis of stereo audio up to 95 seconds with strong structural consistency. InfiniteAudio~\cite{junginfiniteaudio} further improves scalability by introducing a lightweight inference strategy that extends generation to infinite lengths without additional training.

\section{Method}

\subsection{Preliminaries on the Base T2A Model}

DiT-based~\cite{peebles2023scalable,li2024qa,lee2024etta} T2A models are typically trained to learn the reverse of a diffusion process~\cite{ho2020denoising}, progressively removing noise from an initial random state conditioned on a text prompt over multiple diffusion steps. This framework generally consists of three main modules: 1) an audio VAE, responsible for transforming the waveform into a compressed latent representation and reconstructing the waveform from these latents; 2) a pretrained text encoder, which encodes a text prompt into conditioning embeddings; and 3) a Transformer-based diffusion model (DiT), which predicts the denoised audio latents conditioned on the text embeddings. Recent advancements~\cite{evans2024fast,evans2024long,evans2025stable} further enhance the controllability of DiT-based T2A models by incorporating timing embeddings for target durations, enabling variable-length generation.

Building on the DiT-based T2A framework, we select this pretrained T2A model as our base model and aim to adapt it for long-form generation with precise timing control using a training-free approach. Specifically, we develop a base model with the Flan-T5~\cite{chung2024scaling} as the text encoder, trained on audio sampled at 16 kHz mono. The Flan-T5 encoder is frozen during training and its embeddings are injected via cross-attention to condition the diffusion process. We adopt a variable-length training strategy with continuous timing values up to 10 seconds to simulate diverse audio lengths. 

\subsection{Problem Formulation}

We present a training-free method to adapt a pretrained DiT-based T2A model for long-form generation with timing control. We start with a base model, denoted as $f$, which takes a text prompt $y^c$ (i.e., overall audio caption) and a target duration $M$ as input to generate an audio waveform of $M$ seconds, where $M \in (0, M_{\text{max}})$. In practice, $M_{\text{max}}$ is typically set to 10 seconds for most existing T2A models. To enable timing control, our method incorporates an additional input: timing prompts $y^t = \{(p_i^c, s_i, e_i)\}_{i=1}^n$. Here, each element $(p_i^c, s_i, e_i)$ specifies that the audio content described by the fine-grained text prompt $p_i^c$ should occur within the time interval $[s_i, e_i]$.

Specifically, for the timing control capability, FreeAudio employs a two-stage planning-and-generation approach: 1) Planning: Leveraging LLM's capabilities, FreeAudio first generates a detailed timing plan $y^p = \{(p_j^c, t_j, t_{j+1})\}_{j=1}^{k}$, using the overall text prompt $y^c$ and the timing prompts $y^t$. This plan $y^p$ divides the total $M$-second duration into $k$ sequential, non-overlapping windows $\{[t_j, t_{j+1}]\}_{j=1}^{k}$ with $0 = t_1 < \cdots < t_{k+1} = M$, each associated with a recaptioned prompt $p_j^c$. 2) Generation: Based on the resulting timing plan, FreeAudio employs the Decoupling \& Aggregating Attention Control to manipulate the attention outputs across the time axis, aligning each prompt $p_j^c$ with its corresponding time window $[t_j, t_{j+1}]$.

Furthermore, this planning-and-generation schema facilitates long-form audio generation, extending beyond the base model's maximum duration. Instead of directly modeling the entire target duration $M' > M_{\text{max}}$, which exceeds the base model's capacity, our approach first leverages the generated timing plan $y^p$ to partition the task into a sequence of manageable segments, each typically having a duration less than or equal to $M_{\text{max}}$. For smoother transitions between these segments, FreeAudio introduces the Contextual Latent Composition applied at each diffusion sampling step. Additionally, FreeAudio employs the proposed Reference Guidance to facilitate the maintenance of global consistency across the full audio sequence of duration $M’$.

\begin{figure}[t]
\centerline{\includegraphics[width=14.5cm]{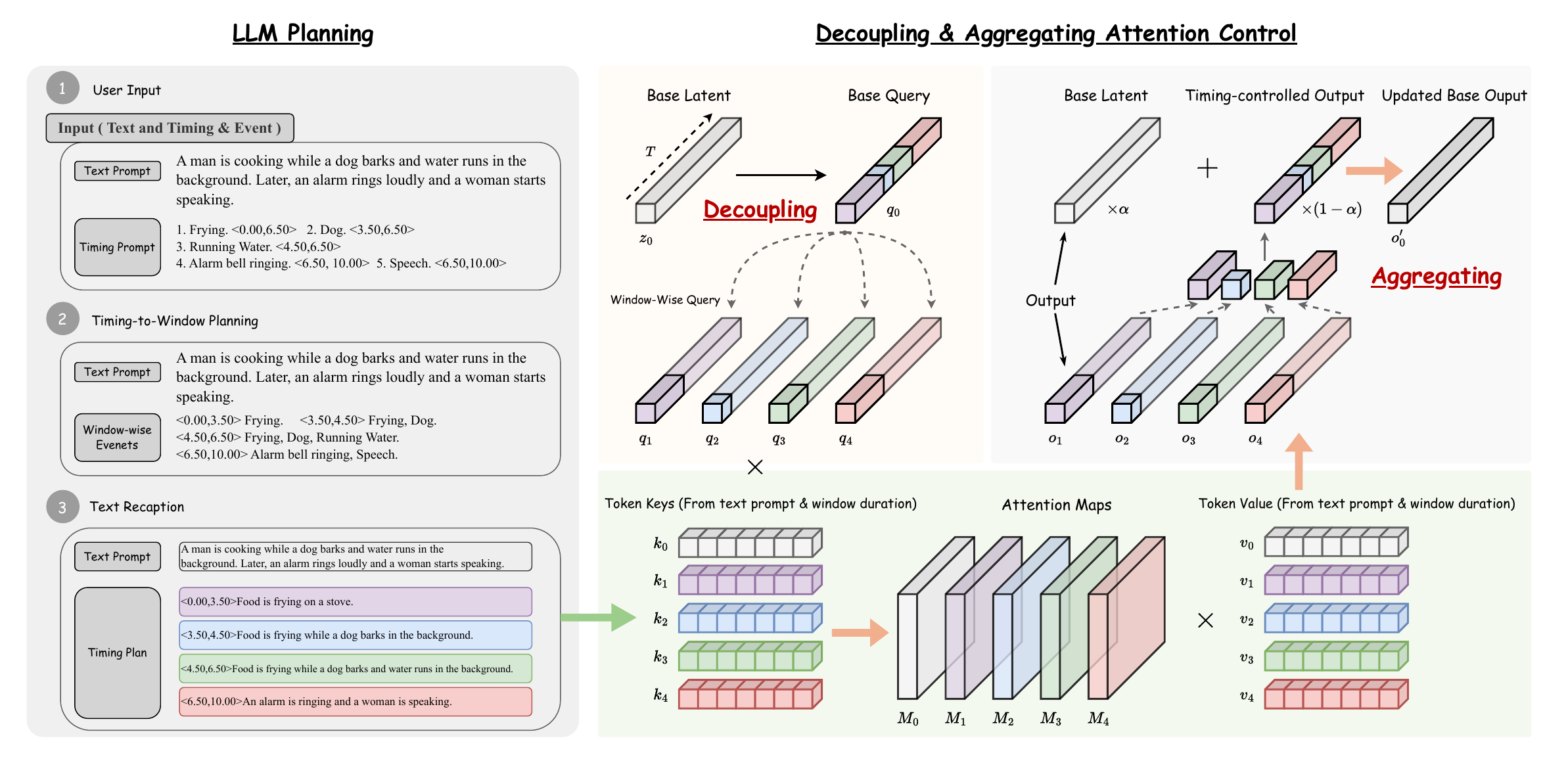}}
\caption{\textit{Left}: Planning Stage, where the LLM parses the text prompt and timing prompts into a sequence of non-overlapping time windows, each associated with a recaptioned prompt. \textit{Right}: Generation Stage, where the Decoupling \& Aggregating Attention Control aligns each recaptioned prompt with its corresponding time window, enabling precise timing control in attention layers.}
\label{fig:pipeline}
\end{figure}

\subsection{LLM Planning}

As shown in Figure~\ref{fig:pipeline} (left), the LLM planning stage takes the text prompt $y^c$ and timing prompts $y^t$ to generate a detailed timing plan $y^p$. This plan consists of a sequence of non-overlapping time windows $[t_j, t_{j+1}]$, each associated with recaptioned prompts $p_j^c$. Specifically, we structure this planning stage using chain-of-thought (CoT)~\cite{sun2024both,wei2022chain} reasoning in two steps: timing-to-window planning to establish the time windows, followed by text recaption to assign a detailed prompt to each window.

\paragraph{Timing-to-Window Planning.} 

The timing-to-window planning stage is necessary for two primary reasons: 1) Temporal Overlaps: Input intervals $[s_i, e_i]$ in the timing prompts $y^t$ can overlap. Merging the text encoder's outputs for separate overlapping prompts (e.g., $p_x^c$ and $p_y^c$) poses challenges for accurate generation guidance. In comparison, establishing non-overlapping windows shifts the fusion to the text input stage (e.g., concatenating prompts such as $p_x^c$ and $p_y^c$), which better leverages the pretrained text encoder's capability for interpreting unified textual descriptions of combined events. 2) Temporal Gaps: The input timing prompts $y^t$ might not cover the entire duration, leaving gaps where no audio event is specified. By leveraging the LLM's planning ability, this stage infers and assigns appropriate audio events for the uncovered periods within the generated timing plan $y^p$. This ensures temporal coherence and semantic consistency in the audio narrative, informed by the context of the planned timing windows and the overall text prompt $y^c$.

Specifically, the timing-to-window planning process involves several steps: First, we define the window boundaries by collecting all unique start ($s_i$) and end ($e_i$) timestamps from $y^t$, inserting $t_1=0$ and $t_{k+1}=M$, and sorting them to obtain $0 = t_1 < t_2 < \dots < t_{k+1} = M$. This defines the non-overlapping windows $[t_j, t_{j+1}]$. Second, we initialize a preliminary description for the window-level events by aggregating all fine-grained text prompts $p_i^c$ whose original intervals $[s_i, e_i]$ encompass the current window $[t_j, t_{j+1}]$. Third, leveraging the overall text prompt $y^c$ and the current window-level event assignments, the LLM performs a planning step to systematically assign the remaining audio events from $y^c$ and distribute them into appropriate window slots, prioritizing the filling of uncovered intervals, thereby finalizing the overall event plan.

\paragraph{Text Recaption.}

This stage focuses on transforming potentially raw or concatenated window-wise event descriptions into detailed and natural prompts $p_j^c$ within the timing plan $y^p$, guided by the text prompt $y^c$. For instance, as illustrated in Figure~\ref{fig:pipeline} (left), a simple list like ``frying, dog, running water'' might be transformed by the LLM into a natural prompt such as ``Food is frying while a dog barks and water runs in the background,'' clarifying the relationships between audio events. This refinement is necessary because the resulting natural language prompts $p_j^c$ align better with both the text prompt $y^c$ and the data distribution the base model was trained on, thereby facilitating higher-quality generation.

\subsection{Timing-Controlled Generation} \label{sec_33}

As illustrated in Figure~\ref{fig:pipeline} (right), FreeAudio employs Decoupling \& Aggregating Attention Control to achieve precise timing control. This subsection begins with a brief review of the standard attention blocks in the underlying DiT-based T2A model (Section~\ref{sec:attn}), followed by an overview of the overall pipeline (Section~\ref{sec:pipeline}). We then introduce the proposed Decoupling \& Aggregating Attention Control in Section~\ref{sec_daac}, which integrates both decoupling and aggregating mechanisms to enhance timing precision and overall consistency.

\paragraph{Attention Blocks in DiT Models.}
\label{sec:attn}
An attention block computes a set of outputs, packed into a matrix ($O$), by mapping a set of queries ($Q$) against a set of key ($K$)-value ($V$) pairs~\cite{vaswani2017attention}. Specifically, each query output is a weighted sum of the values, with weights determined by the similarity (e.g., scaled dot-product) between the query and all keys. This computation can be formalized as:
\begin{equation}
O = \text{Attention}(Q, K, V) = \text{softmax}\left(\frac{QK^T}{\sqrt{d}}\right)V,
\end{equation}
\noindent where $d$ is the dimensionality of the queries and keys. The DiT-based T2A model employs both self-attention and cross-attention blocks. In self-attention, queries, keys, and values are derived from the audio latents. In cross-attention, the queries come from the audio latents, while the keys and values are derived from the text prompt latents.

\paragraph{Overall Pipeline.}
\label{sec:pipeline}

A naive approach to timing control might independently generate each $(t_{j+1}-t_j)$-second audio segment using only its associated prompt $p_j^c$ from the timing plan $y^p$~\cite{wang2024audio}. However, simply concatenating these segments often results in poor inter-window consistency and lacks global guidance from the overall prompt $y^c$. To address these limitations, we construct a batch inference pipeline that jointly leverages both $y^c$ and $y^p$. Specifically, we form a batch of size $1 + k$ (one base latent and $k$ sub-latents) for the base diffusion model, conditioned on the text prompts $[y^c, p_1^c, \dots, p_k^c]$ and the corresponding target durations $[M, (t_2 - t_1), \dots, (t_{k+1} - t_k)]$. The denoised output associated with the base latent (guided by $y^c$ and $M$) serves as the final timing-controlled audio, while the $k$ denoised sub-latents (guided by $p_j^c$ and window durations) are discarded. 

To facilitate information exchange within the batch during denoising, we introduce two attention control mechanisms. First, Decoupling Attention Control segments the base latent according to the timing windows and guides each sub-segment to perform cross-attention independently with its corresponding recaptioned prompt. This design enhances intra-window consistency by allowing finer control within each time window. Second, Aggregating Attention Control concatenates the attended sub-segments based on the timing plan and integrates them with the base latent, enabling global information fusion and ensuring overall semantic and temporal coherence. Aggregating Attention Control is applied in both self-attention and cross-attention blocks, whereas Decoupling Attention Control is applied only in cross-attention blocks, based on our empirical observation that applying Decoupling Attention Control to self-attention adversely affects the precision of timing alignment.

\paragraph{Decoupling \& Aggregating Attention Control.}
\label{sec_daac}
Decoupling Attention Control modifies the query representations in text-audio cross-attention. It segments the base query latents ($q_{\text{base}}$) into window-wise queries $\{q_j\}_{j=1}^k$ according to the timing windows $[t_j, t_{j+1}]$. To promote intra-window consistency via base-to-sub information flow, for each $j$-th sub-latent, we replace its first $(t_{j+1} - t_j)$ positions with $q_j$, while keeping the other positions unchanged. Note that this targeted replacement at the sequence start aligns with the base model's convention for handling variable-length generation.

Aggregating Attention Control enables sub-to-base information flow to enforce precise timing alignment. It operates on the attention outputs from both text-audio cross-attention and audio-only self-attention blocks. Specifically, for each sub-latent, it extracts the active segment $o_j^a$, corresponding to the initial $(t_{j+1}-t_j)$-second positions that match the duration of the $j$-th time window. These active segments are then temporally concatenated following the sequence defined by the timing plan $[t_j, t_{j+1}]$, resulting in a composite timing-controlled output $o_{\text{timing}}$:
% $$o_{\text{timing}} = \text{Concat}\left(o_1^a, o_2^a, \ldots, o_k^a \right)$$
\begin{equation}
o_{\text{timing}} = \text{Concat}\left(o_1^a, o_2^a, \ldots, o_k^a \right).
\end{equation}
Finally, the attention output of the base latent, $o_{\text{base}}$, is updated by linearly interpolating it with $o_{\text{timing}}$:
\begin{equation}
\ o_{\text{base}} = \alpha \cdot o_{\text{base}} + (1 - \alpha) \cdot o_{\text{timing}},
\end{equation}
\noindent where $\alpha$ is a hyper-parameter controlling the fusion ratio.

\begin{figure}[t]
\centerline{\includegraphics[width=14.8cm]{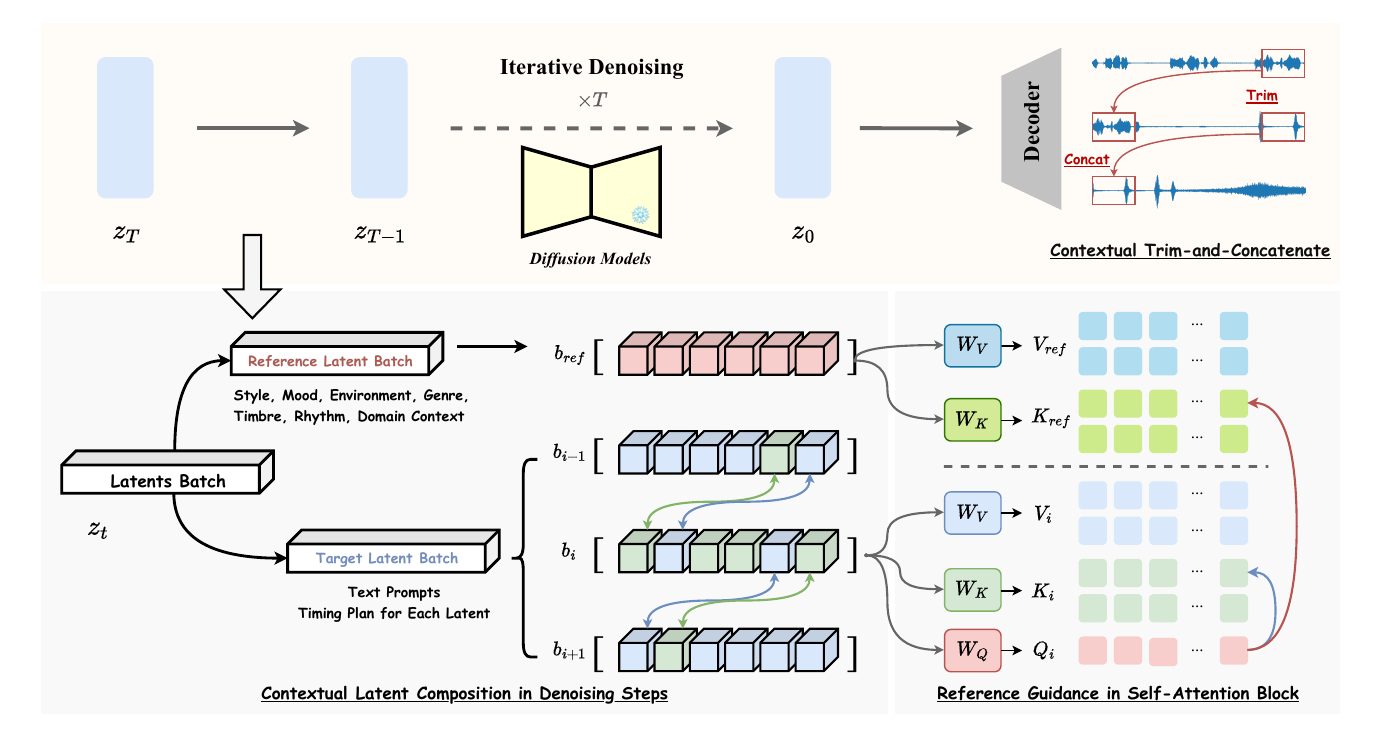}}
\caption{
\textit{Top}: Diffusion denoising process, where latents are progressively denoised and decoded into waveforms, followed by Contextual Cut-and-Concatenate to produce the final long-form audio.
\textit{Bottom Left}: Contextual Latent Composition, where overlapping regions between contextual latents are partially replaced to ensure smooth boundary transitions.
\textit{Bottom Right}: Reference Guidance in the self-attention block, where global consistency is preserved using features from reference latents.
}
\label{fig:long-form}
\end{figure}

\subsection{Long-Form Generation} \label{sec_34}

Generating arbitrarily long audio sequences ($M' > M_{\text{max}}$) presents a challenge for pretrained diffusion models limited by a relatively small inference length ($M_{\text{max}}$). To this end, our proposed planning-and-generation schema addresses this by decomposing the long-form task into generating a sequence of manageable segments using the timing plan $y^p$. While effective, this decomposition inherently leads to three issues: (1) potential discontinuities at the boundaries between generated segments; (2) difficulty in maintaining global consistency across the entire $M'$ duration; and (3) redundant overlap regions in the generated waveform segments. FreeAudio integrates three key techniques to mitigate these issues: (1) Contextual Latent Composition (Section~\ref{sec:clc}), which operates across segments within a single batch during denoising steps to ensure smoother boundary transitions; (2) Reference Guidance (Section~\ref{sec:rg}), which injects global context into the self-attention modules to enhance semantic and stylistic coherence across the full sequence; and (3) Contextual Trim-and-Concatenate (Section~\ref{sec:ccc}), which truncates overlapping regions and concatenates the decoded segments to produce the final long-form audio output.

\paragraph{Contextual Latent Composition.}
\label{sec:clc}

For long-form generation ($M' > M_{\text{max}}$), FreeAudio employs Contextual Latent Composition, which manipulates the predicted denoised outputs during each diffusion step to ensure smooth transitions between segments. The total duration $M'$ is divided into overlapping segments, each typically of length $M_{\text{max}}$, with an overlap of $\epsilon$ seconds between adjacent segments. Each segment is assigned a corresponding text prompt based on the timing plan $y^p$. During generation, Contextual Latent Composition specifically addresses the $\epsilon$-second overlap regions. As illustrated in Figure~\ref{fig:long-form} (bottom left), for any latent segment $b_i$, the $\epsilon$-second overlap with its preceding segment $b_{i-1}$ is handled by retaining the first $\epsilon/2$ seconds of overlap from $b_i$ and the last $\epsilon/2$ seconds from $b_{i-1}$. Similarly, for the overlap between $b_i$ and its succeeding segment $b_{i+1}$, the first $\epsilon/2$ seconds are taken from $b_{i+1}$ and the last $\epsilon/2$ seconds from $b_i$. This strategy ensures seamless transitions between adjacent segments by blending their overlapping regions at the latent level. The operation can be formalized as:
% $$
\begin{equation}
\begin{aligned}
b_i[:\epsilon] &\leftarrow \text{Concat}\left(b_i[:\epsilon/2],\ b_{i-1}[-\epsilon/2:]\right), \
b_i[-\epsilon:] &\leftarrow \text{Concat}\left(b_{i+1}[:\epsilon/2],\ b_i[-\epsilon/2:]\right),
\end{aligned}
\end{equation}
% $$
where $\text{Concat}(\cdot)$ denotes concatenation along the temporal axis.

\paragraph{Reference Guidance.}
\label{sec:rg}

To maintain global consistency across the generated long-form audio, FreeAudio proposes Reference Guidance, which operates on the self-attention modules. Specifically, during the denoising process, we utilize features derived from reference audio (either input or synchronously generated) to provide global anchoring. Let $q_i$ be the query and $o_i$ be the original self-attention output for the $i$-th segment. Let $k_{\text{ref}}$ and $v_{\text{ref}}$ be the key and value pairs derived from the reference audio features. The final self-attention output $o_i'$ is given by:
\begin{equation}
o_i' = \lambda \ast \text{Attention}(q_i, k_{\text{ref}}, v_{\text{ref}}) + (1 - \lambda) \ast o_i,
\end{equation}
\noindent where $\lambda$ is a hyper-parameter controlling the strength of the guidance from the reference audio.

\paragraph{Contextual Trim-and-Concatenate.}
\label{sec:ccc}

After the diffusion denoising process, the latent representations are passed through a pretrained VAE decoder to reconstruct the corresponding waveform segments. Due to the overlapping strategy employed in Contextual Latent Composition, adjacent segments share an overlap region of $\epsilon$ seconds to ensure smooth transitions during generation. To obtain the final long-form audio, we trim the overlapping regions by retaining a single copy for each pair of adjacent decoded segments, avoiding duplication, and then sequentially concatenate the segments. This ensures that redundant artifacts in the overlap regions are eliminated, resulting in a temporally coherent and acoustically smooth long-form audio output.

\begin{table}[t]
\centering
\caption{Objective and subjective evaluation results on the AudioCondition test set. For each metric, the best result is in bold and the second-best is underlined. * denotes models trained by ourselves.\\ $^{\ddagger}$ denotes models evaluated under a different SED model.}
\label{tab1}
\begin{tabular}{cccccccc}
\toprule
\multirow[c]{2}{*}[-0.5ex]{Method}  & \multicolumn{2}{c}{\textbf{Temporal (Obj.)}} & \multicolumn{3}{c}{\textbf{Generation (Obj.)}} & \multicolumn{2}{c}{\textbf{Subjective}} \\
\cmidrule(lr){2-3} \cmidrule(lr){4-6} \cmidrule(lr){7-8}
 & Eb↑ & At↑ & FAD↓ & KL↓ & CLAP↑ & Temporal↑ & Ovl↑ \\
\midrule
Ground Truth         & 43.37 & 67.53 & -    & -    & 0.377 & 4.53 & 4.20 \\
\midrule
AudioLDM Large       & 6.79  & 35.66 & 3.95 & 2.46 & 0.260 & 1.98 & 3.01 \\
AudioLDM 2 Full      & 7.75  & 42.41 & 3.07 & 1.92 & 0.279 & -    & -    \\
AudioLDM 2 Full Large& 6.93  & 20.47 & 3.68 & 2.15 & 0.283 & -    & -    \\
Tango                & 1.60  & 26.51 & 2.82 & 1.93 & 0.245 & 2.53 & 3.01 \\
Stable Audio~*       & 11.28 & 51.67 & \underline{1.93} & \underline{1.75} & \underline{0.318} & \underline{2.63} & \underline{3.30} \\
\midrule
CCTA                 & 14.57 & 18.27 & -    & -    & -     & -    & -    \\
MC-Diffusion         & 29.07 & 47.11 & -    & -    & -     & -    & -    \\
Tango + LControl     & 21.46 & 55.15 & -    & -    & -     & -    & -    \\
AudioComposer-S      & 43.51 & 60.83 & -    & -    & -     & -    & -    \\
AudioComposer-L      & \textbf{44.40} & \underline{63.30} & - & - & - & - & - \\
\midrule
TG-Diff$~^{\ddagger}$ & 26.70 & 60.06 & 2.66 & -    & 0.244 & -    & -    \\
FreeAudio            & \underline{44.34} & \textbf{68.50} & \textbf{1.92} & \textbf{1.73} & \textbf{0.321} & \textbf{4.10} & \textbf{3.65} \\
\bottomrule
\end{tabular}
\small
\end{table}

\section{Experiments}

\subsection{Timing-Controlled Audio Generation}

\paragraph{Experiment Settings.}

To objectively evaluate the controllability and scalability of the proposed method, we utilize the publicly available test split from AudioCondition~\cite{guo2024audio}, which consists of 1,110 audio samples from AudioSet-Strong~\cite{hershey2021benefit}. Each sample is annotated with frame-level timestamps for sound events, covering a vocabulary of 456 event classes. The fine-grained temporal annotations make this dataset particularly well-suited for evaluating timing-controllable generation.

\paragraph{Evaluation Metrics.}

For evaluating timing control, we follow prior work~\cite{guo2024audio,wang2025audiocomposer} on temporal order control and employ two timing metrics: the event-based measures (Eb) and the clip-level macro F1 score (At), computed using a sound event detection (SED) system~\cite{mesaros2016metrics}. We conduct evaluations using the open-source PB-SED system~\cite{ebbers2021forward,ebbers2021self,ebbers2022pre}. Notably, previous methods~\cite{guo2024audio,wang2025audiocomposer} for timing-controllable audio generation only reported timing-related metrics while neglecting audio generation quality. To address this gap, we additionally assess the audio quality of the generated samples based on standard metrics: Fréchet Audio Distance (FAD), Kullback–Leibler (KL) divergence~\cite{liu2023audioldm}, and CLAP score~\cite{wu2023large}. Furthermore, to complement the objective evaluation, we introduce two subjective measures: temporal alignment and overall audio quality. 

\paragraph{Main Results.}

\begin{wrapfigure}{r}{0.6\textwidth}
\vspace{-7.5mm}
\centering
\includegraphics[width=0.58\textwidth]{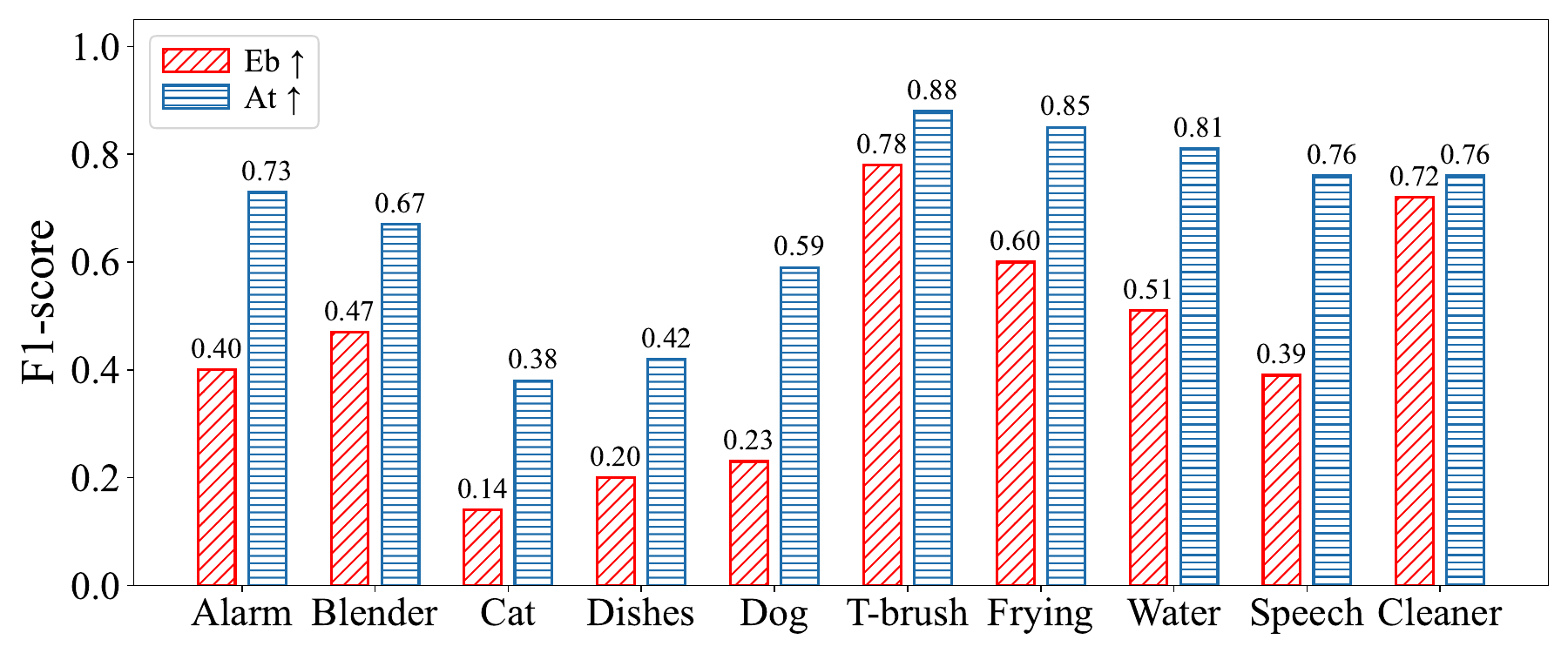}
\caption{Event-wise Eb and At scores across 10 categories of FreeAudio on the AudioCondition test set.}
\label{fig4}
\vspace{-5mm}
\end{wrapfigure}

We compare FreeAudio with several state-of-the-art T2A models, including AudioLDM, AudioLDM 2, Tango, and our in-house implementation of Stable Audio. Additionally, we include models that incorporate temporal conditioning signals, such as MC-Diffusion~\cite{guo2024audio} and AudioComposer~\cite{wang2025audiocomposer}, as well as the training-free baseline TG-Diff~\cite{du2024controllable}. Notably, TG-Diff reports both timing and audio quality metrics using a training-free framework but adopts a different sound event detection model~\cite{turpault2019sound} compared to the other baselines. CCTA (Control-Condition-to-Audio) is a baseline from MC-Diffusion that uses only control conditions without textual input. Tango + LControl is a variant from AudioComposer, built upon Tango with language-based control. As shown in Table~\ref{tab1}, FreeAudio achieves superior overall performance by effectively integrating timing control with high-quality T2A generation. Furthermore, Figure~\ref{fig4} presents the category-wise Eb and At scores on the AudioCondition test set, providing a more fine-grained evaluation of FreeAudio’s performance across different sound events. The results indicate that FreeAudio delivers strong temporal alignment and segment-level quality for events with stable acoustic patterns and longer durations, such as \textit{Toothbrush}, \textit{Frying}, and \textit{Vacuum Cleaner}. In contrast, for events like \textit{Cat}, \textit{Dog}, and \textit{Dishes}—which are typically shorter, acoustically weaker, or exhibit ambiguous label boundaries—FreeAudio demonstrates relatively lower alignment accuracy but maintains consistent segment-level performance. These results demonstrate the adaptability and robustness of FreeAudio across diverse sound events.

To validate the effectiveness of FreeAudio, we summarize both objective and subjective evaluation results in Table~\ref{tab1}. FreeAudio consistently achieves strong performance across timing controllability and audio generation quality metrics, outperforming existing baselines. In particular, it demonstrates reliable temporal alignment and maintains high perceptual audio quality in subjective evaluations, confirming its effectiveness for controllable and high-fidelity T2A generation.

\paragraph{Ablation Study.}

\begin{wrapfigure}{r}{0.6\textwidth}
\vspace{-8.5mm}
\centering
\includegraphics[width=0.58\textwidth]{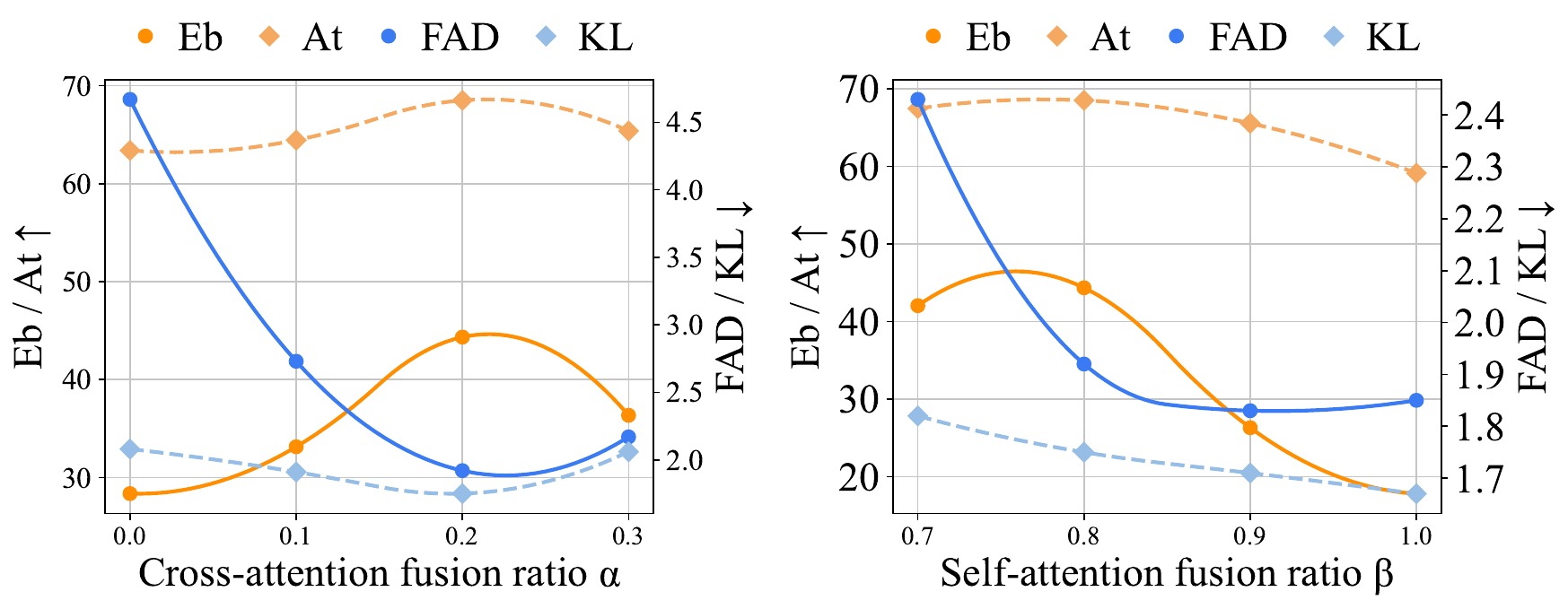}
\vspace{0.5mm}
\caption{Ablation study on the effect of cross-attention fusion ratio $\alpha$ and self-attention fusion ratio $\beta$ on generation performance. Left: $\alpha = 0.2$, varying $\beta$. Right: $\beta = 0.8$, varying $\alpha$. Higher $\beta$ improves temporal alignment, while balanced $\alpha$ enhances overall audio quality.}
\label{fig5}
\vspace{-4mm}
\end{wrapfigure}

To investigate the influence of attention fusion mechanisms on generation performance, we conduct an ablation study focusing on two key hyperparameters: the cross-attention fusion ratio ($\alpha$) and the self-attention fusion ratio ($\beta$). Here, $\alpha$ controls the contribution of cross-attention guided by the base prompt, while $\beta$ governs the strength of self-attention within each audio segment. Adjusting these ratios affects the balance between timing controllability and audio quality. Generally, reducing $\alpha$ and increasing $\beta$ enhances temporal alignment by strengthening local dependencies, but may compromise overall audio fidelity. Conversely, a higher $\alpha$ leverages the base prompt to maintain global coherence, and a balanced $\beta$ leads to smoother transitions across time windows, resulting in higher-quality audio. To systematically evaluate their impact, we perform two sets of controlled experiments: (1) fixing $\alpha$ at 0.2 while varying $\beta$ to study its effect on timing controllability and generation quality; and (2) fixing $\beta$ at 0.8 while varying $\alpha$ to examine its influence on overall performance. The results, as shown in Figure~\ref{fig5}, demonstrate that both $\alpha$ and $\beta$ are critical for achieving an optimal trade-off between precise timing control and high-fidelity audio generation.

\subsection{Long-Form Audio Generation}

\paragraph{Experiment Settings.}

We adopt the standard AudioCaps~\cite{kim2019audiocaps} and MusicCaps~\cite{agostinelli2023musiclm} datasets as evaluation benchmarks. The AudioCaps test set consists of 979 audio clips from YouTube, while MusicCaps comprises 5521 music clips, also sourced from YouTube. Since both datasets provide 10-second audio segments, we follow the evaluation protocol of Stable Audio and extend it to support variable-length generation at a 16kHz sampling rate. Notably, the provided captions correspond only to the original 10-second segments and do not offer consistent semantic coverage over longer durations. Therefore, in our experiments, the 10-second clips are used as references, while the generated audio samples are extended to variable lengths ranging from 10 to 90 seconds.

\begin{table}[t]
\begin{center}
\caption{Objective evaluation on AudioCaps and MusicCaps test sets at different output lengths. AG / MG denote AudioGen and MusicGen, respectively. For Stable Audio, we use the open-source version, Stable Audio Open. $^{\dagger}$ indicates results directly taken from reported metrics in existing literature. Other results are evaluated using publicly available checkpoints.} \label{tab:longform_results}
\begin{tabular}{cccccccccc}
\toprule
\multirow[c]{2}{*}[-0.5ex]{Method} & \multirow[c]{2}{*}[-0.5ex]{\makecell{Output\\Length}} & \multicolumn{4}{c}{AudioCaps} & \multicolumn{4}{c}{MusicCaps} \\
\cmidrule(lr){3-6} \cmidrule(lr){7-10}
& & FAD↓ & KL↓ & FD↓ & CLAP↑ & FAD↓ & KL↓ & IS↑ & CLAP↑ \\
\midrule
AudioLDM                  & 10 sec & 4.96 & 2.17 & 29.29 & 0.373     & 3.51 & 1.63 & \textbf{3.31}  & 0.304 \\
AG$~^{\dagger}$ / MG$~^{\dagger}$     & 10 sec & 1.82 & 1.69 & -     &  -    & 3.40 & 1.23 & -   & 0.320 \\
AudioLDM 2$~^{\dagger}$   & 10 sec & 1.78 & 1.60 & -     & 0.191 & 3.13 & \textbf{1.20} & -    & 0.301 \\
Stable Audio$~^{\dagger}$        & 10 sec & 3.60 & 2.32 & 38.27 & 0.340 & 3.51 & 1.32 & 2.93 & \textbf{0.480}  \\
FreeAudio                 & 10 sec & \textbf{1.52} & \textbf{1.51} & \textbf{18.30} & \textbf{0.538} & \textbf{2.36} & 1.68 & 3.10 & 0.368  \\
\midrule
AudioLDM      & 26 sec & 5.36 & 2.23 & 30.73 & 0.371 & 3.98 & 1.73 & \textbf{2.88} & 0.294  \\
AG / MG       & 26 sec & 2.98 & 1.65 & 17.45 & 0.300 & 5.28 & 1.67 & 2.02 & \textbf{0.371} \\
AudioLDM 2    & 26 sec & 2.02 & 1.77 & 17.32 & 0.312 & 4.26 & \textbf{1.45} & 2.39 & 0.352 \\
Stable Audio  & 26 sec & 4.23 & 2.27 & 40.38 & 0.335 & 3.45 & 1.54 & 2.06 & 0.310 \\
FreeAudio     & 26 sec & \textbf{1.85} & \textbf{1.49} & \textbf{16.21} & \textbf{0.527} & \textbf{2.46} & 1.65 & 2.17 & 0.359 \\
\midrule
AudioLDM      & 90 sec & 5.45 & 2.35 & 33.75 & 0.362  & 4.19 & 1.78 & \textbf{2.77} & 0.290   \\
AG / MG       & 90 sec & 3.44 & 1.78 & \textbf{18.27} & 0.210 & 6.43 & 1.75 & 2.34 & 0.295 \\
AudioLDM 2    & 90 sec & 2.26 & 1.97 & 22.48 & 0.288 & 4.23 & 1.68 & 2.46 & 0.301 \\
Stable Audio  & 90 sec & 4.35 & 2.48 & 45.27 & 0.345 & 3.37 & \textbf{1.61} & 2.23 & \textbf{0.358} \\
FreeAudio     & 90 sec & \textbf{2.19} & \textbf{1.68} & 19.95 & \textbf{0.495} & \textbf{2.57} & 1.76 & 2.29 & 0.342 \\
\bottomrule
\end{tabular}
\small
\end{center}
\vspace{-2.5mm}
\end{table}

\paragraph{Evaluation Metrics.}

We adopt FAD, KL, Fréchet Distance (FD), Inception Score (IS), and CLAP score as evaluation metrics for long-form audio generation, following the evaluation protocol proposed in Stable Audio. For KL divergence, each generated audio sample is divided into overlapping sub-windows. We compute the logits for each window, average them across all windows, and apply the softmax operation to obtain the final probability distribution. The same strategy is applied to KL evaluation at 16\,kHz. For FAD, FD, and IS, we similarly use the overlapping sub-window scheme and report the mean of the scores as the final evaluation result. 

For the CLAP metric, we compute the cosine similarity between the text embedding of the prompt and the audio embedding of the generated audio. Following Stable Audio, we adopt the Feature Fusion variant of CLAP to accommodate long-form audio. Specifically, three 10-second crops are extracted from the beginning, middle, and end of each audio sample and fused into a global representation.

\paragraph{Main Results.}

We evaluate the long-form generation performance of FreeAudio on the AudioCaps and MusicCaps datasets to assess its adaptability and stability across varying audio durations. Specifically, we conduct experiments at three target lengths: 10 seconds, 26 seconds, and 90 seconds, as summarized in Table~\ref{tab:longform_results}. For the AudioCaps benchmark, FreeAudio is compared against several representative models, including AudioLDM, AudioGen~\cite{kreuk2022audiogen}, AudioLDM2 and Stable Audio. For MusicCaps, we include AudioLDM, MusicGen~\cite{copet2023simple}, AudioLDM2 and Stable Audio as baselines. All models are evaluated under a unified preprocessing and evaluation protocol. For 10-second generation, $^{\dagger}$ indicates that the results are directly taken from reported metrics in existing literature; all other models are evaluated using publicly released checkpoints. The results demonstrate that FreeAudio consistently maintains stable generation quality across different durations. Notably, it achieves high audio fidelity and semantic alignment even at extended lengths such as 26 and 90 seconds, highlighting its effectiveness and robustness for long-form generation.

To comprehensively evaluate the performance of FreeAudio in long-form audio generation, we conducted a subjective study on the AudioCaps and MusicCaps datasets. Beyond standard audio quality, two additional dimensions—coherence and style consistency—were introduced to assess the model’s ability to maintain semantic continuity and overall stylistic integrity across long audio sequences. Participants rated the generated samples along these three dimensions using a five-point Likert scale, where higher scores indicate better perceptual quality. As shown in Table~\ref{tab3_sub}, FreeAudio demonstrates consistently strong performance across all dimensions. These results suggest that FreeAudio, even without additional training, can generate long-form audio that is both structurally coherent and stylistically consistent.

\begin{table}[htbp]
\begin{center}
% \vspace{-2.5mm}
\caption{Subjective evaluation results for AudioCaps and MusicCaps.}
\begin{tabular}{ccccccc}
\toprule
\multirow{2}{*}{Method} & \multicolumn{3}{c}{AudioCaps} & \multicolumn{3}{c}{MusicCaps} \\
\cmidrule(lr){2-4} \cmidrule(lr){5-7}
& Quality↑ & Consist.↑ & Coheren.↑ & Quality↑ & Consist.↑ & Coheren.↑ \\
\midrule
AudioLDM     & 2.53 & 3.29 & 3.65 & 3.07 & 3.65 & 3.53 \\
AG / MG      & 2.80 & 3.48 & 3.72 & 3.28 & 3.88 & 3.64 \\
AudioLDM 2   & 2.84 & 3.68 & 3.80 & 3.48 & 4.02 & 4.06 \\
Stable Audio & 3.32 & 3.76 & 3.92 & 3.52 & 4.06 & 4.28 \\
FreeAudio    & \textbf{4.08} & \textbf{4.12} & \textbf{4.20} & \textbf{3.88} & \textbf{4.36} & \textbf{4.34} \\
\bottomrule
\end{tabular}
\label{tab3_sub}
\end{center}
% \vspace{-5mm}
\end{table}

\begin{figure}[t]
\centerline{\includegraphics[width=14.75cm]{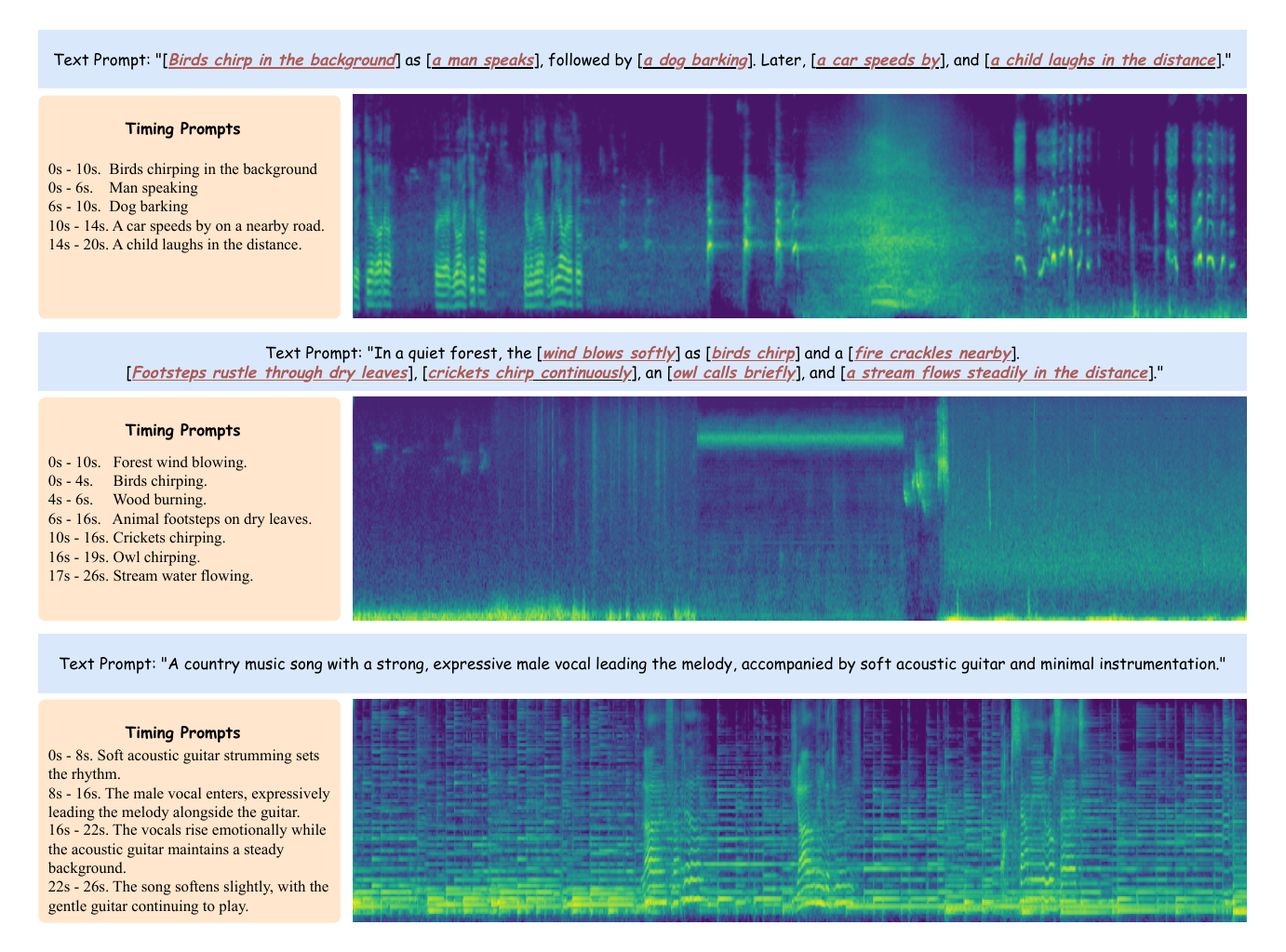}}
\vspace{-2.5mm}
\caption{Spectrograms of timing-controlled long-form audio generations.}
\label{fig:casestudypaper}
\vspace{-7mm}
\end{figure}

\paragraph{Case Studies.}

We present a set of case studies in Figure~\ref{fig:casestudypaper}, showcasing spectrograms of timing-controlled long-form audio samples. These examples span both general audio and music generation, demonstrating the versatility of FreeAudio across different content domains. In all cases, the generated audio closely follows the specified timing prompts, ensuring that events occur at the correct temporal positions. Moreover, the transitions between events are smooth and acoustically coherent, resulting in natural-sounding audio even over extended durations. This highlights FreeAudio’s strength in producing temporally aligned and harmoniously structured outputs across diverse long-form scenarios.

In addition, FreeAudio benefits from a robust T2A backbone trained on diverse audio-text datasets beyond AudioCaps~\cite{kim2019audiocaps} and MusicCaps~\cite{agostinelli2023musiclm} (see Appendix~\ref{appendix_datasets} for details), as well as an LLM planner capable of interpreting noisy or out-of-distribution prompts. The LLM module not only corrects errors such as typos, unconventional syntax, and inconsistent formatting, but also adapts to previously unseen concepts or expressions not present in training data, allowing FreeAudio to operate reliably in real-world scenarios. We provide qualitative examples of the LLM’s robustness in Appendix~\ref{appendix_llm}.

\paragraph{Ablation Study.}

To further validate the effectiveness of Reference Guidance (RG), we perform an ablation study on long-form audio generation. Beyond standard metrics, two additional measures—Intra-LPIPS and Intra-CLAP—are introduced to assess cross-view coherence, computed by measuring the internal similarity between 10-second audio segments extracted via a sliding window. Specifically, Intra-CLAP is the cosine similarity between CLAP embeddings of different segments.

To isolate the contribution of RG, we first generate audio samples under identical conditions without RG and compute their performance across all evaluation metrics as the baseline. We then compare this against several RG-enabled variants with different interpolation coefficients $\lambda$. All evaluations are conducted on 26-second samples. Experimental results, summarized in Table~\ref{tab4}, show that incorporating RG significantly reduces distributional divergence between local segments. Moreover, as $\lambda$ increases, we observe a consistent improvement in intra-audio coherence, reflected by lower I-LPIPS and higher I-CLAP scores. These results demonstrate that Reference Guidance is crucial for producing coherent and stylistically consistent long-form audio, particularly in maintaining temporal and semantic continuity over extended durations.

\begin{table}[htbp]
\begin{center}
\vspace{-5mm}
\caption{Effect of varying interpolation coefficient $\lambda$ in Reference Guidance on AudioCaps test set.}
% \vspace{2.5mm}
\label{tab4}
\begin{tabular}{cccccc}
\toprule
\multirow{2}{*}{Setting} & \multicolumn{5}{c}{Metrics} \\
\cmidrule(lr){2-6}
& FAD↓ & KL↓ & FD↓ & Intra-LPIPS↓ & Intra-CLAP↑ \\
\midrule
Reference       & 0.25 & 0.64 & 6.94 & 0.82 & 0.763 \\
\midrule
$\lambda = 0.0$ & 0.27 & 0.59 & 7.29 & 0.76 & 0.723  \\
$\lambda = 0.1$ & 0.23 & 0.41 & 6.61 & 0.52 & 0.801  \\
$\lambda = 0.2$ & \textbf{0.21} & \textbf{0.26} & \textbf{5.70} & \textbf{0.36} & \textbf{0.860}  \\
\bottomrule
\end{tabular}
\end{center}
% \vspace{-5mm}
\end{table}

While RG enhances semantic coherence and stylistic consistency across long-form audio, we observe a potential side effect. When the interpolation weight $\lambda$ is set too high, the generated audio tends to exhibit overly uniform vocal characteristics, leading to reduced diversity in local details. Specifically, the audio may sound acoustically homogeneous, with repeated events and diminished variations in timbre, rhythm, and expressiveness, resulting in a somewhat monotonous overall structure. To balance global consistency and local diversity, we recommend setting $\lambda$ within the range of 0.1 to 0.2.

\section{Conclusion}

In this work, we propose FreeAudio, a novel training-free T2A framework that achieves superior performance in timing-conditioned text-to-audio generation. For 10-second timing-conditioned generation tasks, FreeAudio surpasses previous training-free methods by a large margin and achieves results comparable to state-of-the-art training-based models, without requiring any temporally aligned audio-text data for training. Furthermore, for long-form audio generation, FreeAudio makes the first attempt to enable timing-conditioned long-form synthesis, delivering performance on par with training-based approaches while significantly reducing computational overhead.

\bibliography{neurips_2025}
\bibliographystyle{tmlr}

\clearpage

\appendix

\section{Experimental Details}

\subsection{Datasets}\label{appendix_datasets}

Although the proposed method is entirely training-free with respect to the timing control and long-form generation modules, the underlying text-to-audio generation backbone (T2A model) still requires pretraining. To ensure reproducibility and experimental completeness, we provide detailed information about all datasets used for training and evaluation in the supplementary materials, as summarized in Table~5. We pretrain our T2A model using several publicly available text-audio aligned datasets. All training samples are standardized to 16kHz sampling rate, mono channel format, and constrained to a maximum duration of 10 seconds to align with the training paradigm of diffusion models.

\begin{table}[htbp]
\centering
\caption{Details about audio-text datasets we use.}
\label{tab_app_1}
\begin{tabular}{cccc}
\toprule
\textbf{Dataset} & \textbf{Hours}(h)  & \textbf{Number}  & \textbf{Text}  \\
\midrule
AudioCaps  & 109   & 44K  & caption \\
AudioSet   & 5800  & 2M   & label\\
WavCaps    & 7090  & 400K & caption\\
FSD50k     & 108   & 51K  & label\\
VGG-Sound  & 550   & 210k & label\\
\hline
MTT        & 200   & 24K  & caption\\
MSD        & 7333  & 880K & caption\\
FMA        & 900   & 11K  & caption\\
\bottomrule
\end{tabular}
\end{table}

\subsection{Model Configurations}

Our diffusion model is built upon the DiT (Diffusion Transformer) architecture, following a latent diffusion modeling (LDM) paradigm that offers strong generative capabilities and effective context modeling. The model is conditioned on three types of inputs: a natural language prompt (\texttt{prompt}), the start time of the audio segment (\texttt{seconds\_start}, which is set to 0 by default in our task), and the total duration (\texttt{seconds\_total}). All conditions are embedded into a 768-dimensional feature space. The prompt is encoded using a pretrained Flan-T5-Base model, while \texttt{seconds\_start} and \texttt{seconds\_total} are treated as numerical inputs with value ranges restricted between 0 and 10 seconds.

The backbone of the diffusion network adopts a DiT structure with 24 layers and 24 attention heads, each with an embedding dimension of 1536~\cite{evans2024long}. The model supports both cross-attention and global conditioning: cross-attention is applied to all types of conditional inputs, while global conditioning specifically handles duration-related control signals. The internal token dimension of the diffusion model is set to 64, with a conditional token dimension of 768 and a global condition embedding dimension of 1536. The generated latent representation has the same dimensionality as \texttt{io\_channels}, which is 64.

\subsection{Compression Networks}

To train the audio autoencoder, we adopt a variational autoencoder (VAE) architecture based on the Oobleck framework at a sampling rate of 16kHz~\cite{evans2024long}. The model is trained from scratch on large-scale publicly available text-audio paired datasets. The encoder and decoder are symmetric, each using a base channel size of 128, with channel multipliers $1, 2, 4, 8, 16$ and strides $2, 2, 4, 4, 10$. The encoder maps the input waveform into a 128-dimensional latent representation, while the decoder reconstructs the waveform from a 64-dimensional latent code. Snake activation is applied throughout the network, and no final tanh activation is used in the decoder. The overall downsampling ratio is 640, and both input and output are mono-channel waveforms. The bottleneck is implemented as a variational layer.

\subsection{Training Strategies}

To improve convergence stability and generation quality, we adopt the following training strategies~\cite{evans2025stable}. We apply Exponential Moving Average (EMA) to the model parameters to smooth updates and mitigate training fluctuations, leading to more coherent and stable generation results. For optimization, we use the AdamW optimizer with a learning rate of $5 \times 10^{-5}$, $\beta_1 = 0.9$, $\beta_2 = 0.999$, and a weight decay of $1 \times 10^{-3}$ to regularize parameter magnitudes and prevent overfitting. For learning rate scheduling, we adopt the InverseLR strategy, where the learning rate at step $t$ is defined as:
\begin{equation}
\eta_t = \eta_0 \times \left(1 + \frac{t}{\gamma}\right)^{-\text{power}}
\end{equation}
with $\gamma = 10^6$ and $\text{power} = 0.5$. We also apply a warm-up ratio of 0.99 to maintain a high learning rate during the early phase of training for faster convergence, followed by gradual decay for improved stability.

\section{Evaluation}

\subsection{Objective Metrics}

We conduct a comprehensive evaluation of our proposed models from the perspectives of audio quality, semantic alignment, and inference efficiency. For objective evaluation, we adopt the following metrics: Fréchet Audio Distance (FAD), Kullback-Leibler (KL) divergence, Inception Score (IS), Fréchet Distance (FD), and the LAION-CLAP score.

FAD, adapted from the Fréchet Inception Distance (FID) in the image domain, measures the distributional difference between generated and reference audio based on VGGish embeddings, and serves as our primary metric for audio fidelity. KL divergence is computed using the PANN tagging model and quantifies the difference in posterior distributions of acoustic events between the generated and ground truth audio, reflecting event-level consistency. IS captures both the diversity and discriminability of the generated samples by computing entropy over predicted event class distributions. FD is structurally similar to FAD but operates in a more general embedding space; it tends to be less stable in audio tasks and is thus included as a supplementary metric.

The CLAP score is defined as the cosine similarity between the CLAP embeddings of the generated audio and its corresponding text caption. Given audio embedding $a$ and text embedding $t$, the score is computed as:
\[
\text{CLAP}(a, t) = \frac{a \cdot t}{\|a\| \, \|t\|}
\]
This metric captures the semantic alignment between audio and text in a shared representation space. All objective metrics are computed using the AudioLDM evaluation toolkit.

\subsection{Subjective Evaluation}

For subjective evaluation, we recruited 20 human raters to assess two distinct tasks: timing-controlled generation and long-form audio generation. In the timing-controlled generation task, we evaluated the model from two perspectives: (i) temporal alignment accuracy, which measures how well the generated audio conforms to the specified timing constraints, and (ii) overall perceptual quality, which reflects the general auditory impression of the output. In the long-form audio generation task, we conducted evaluations on both the AudioCaps and MusicCaps datasets. For each sample, raters scored the results along three dimensions: overall audio quality, temporal coherence across segments, and stylistic consistency throughout the sequence. These criteria are designed to assess the model’s ability to maintain semantic integrity and structural stability in long-duration audio generation.

\section{Case Study}

We provide additional qualitative results and audio samples on our demo page, including various timing-controlled and long-form generation cases: \href{https://freeaudio.github.io/FreeAudio/}{https://freeaudio.github.io/FreeAudio/}.

\section{LLM Planning}

\subsection{LLM Robustness on Unseen or Noisy Prompts}
\label{appendix_llm}

To further assess the robustness of FreeAudio's planning module, we examine its ability to handle noisy and previously unseen prompts. This capability is largely attributed to the use of a LLM that interprets and decomposes text and timing prompts into structured sub-instructions.

For instance, given the following noisy prompt:

\noindent
\textbf{Text prompt:} \textit{A man is cooking while dog bakring \&\& water runs … later ALARM rings loud and she talks.} \\
\textbf{Timing prompts:} \textit{Frying. <0.0,8.0>, Dog bakring. 0s-4s., Running water. <4.0,8.0>, Alarm ringing. From 8 to 10., Woman speaking. 8$\sim$10 sec.}

\medskip

\noindent
The planner successfully produces the following structured timing plan:
\begin{itemize}[leftmargin=*]
    \item \textit{Food is frying while a dog barks. <0.0,4.0>}
    \item \textit{Food is frying while water runs in the background. <4.0,8.0>}
    \item \textit{An alarm rings and a woman is speaking. <8.0,10.0>}
\end{itemize}

This example demonstrates the planner's ability to extract relevant semantic elements and unify inconsistent formats into coherent sub-instructions. By leveraging the generalization power of the LLM, FreeAudio can effectively handle real-world, imperfect user prompts without requiring strict input formatting, enhancing its practical usability in open-ended scenarios.

\subsection{LLM Planning Prompts}

We design a set of prompts to guide GPT-4o in the temporal planning stage of FreeAudio. Given a global audio caption and a set of event-level timing hints, the model generates a structured timeline by first inferring a sequence of non-overlapping time windows and performing event fusion and gap completion when necessary. It then produces a recaptioned prompt for each time window, ensuring that the resulting descriptions align with the input style expected by the downstream generative models. This process highlights the capability of GPT-4o in organizing events, understanding temporal context, and restructuring language, serving as a critical component for generating coherent multi-event audio scenes in FreeAudio.

\begin{figure*}[htbp]
\centerline{\includegraphics[width=18cm]{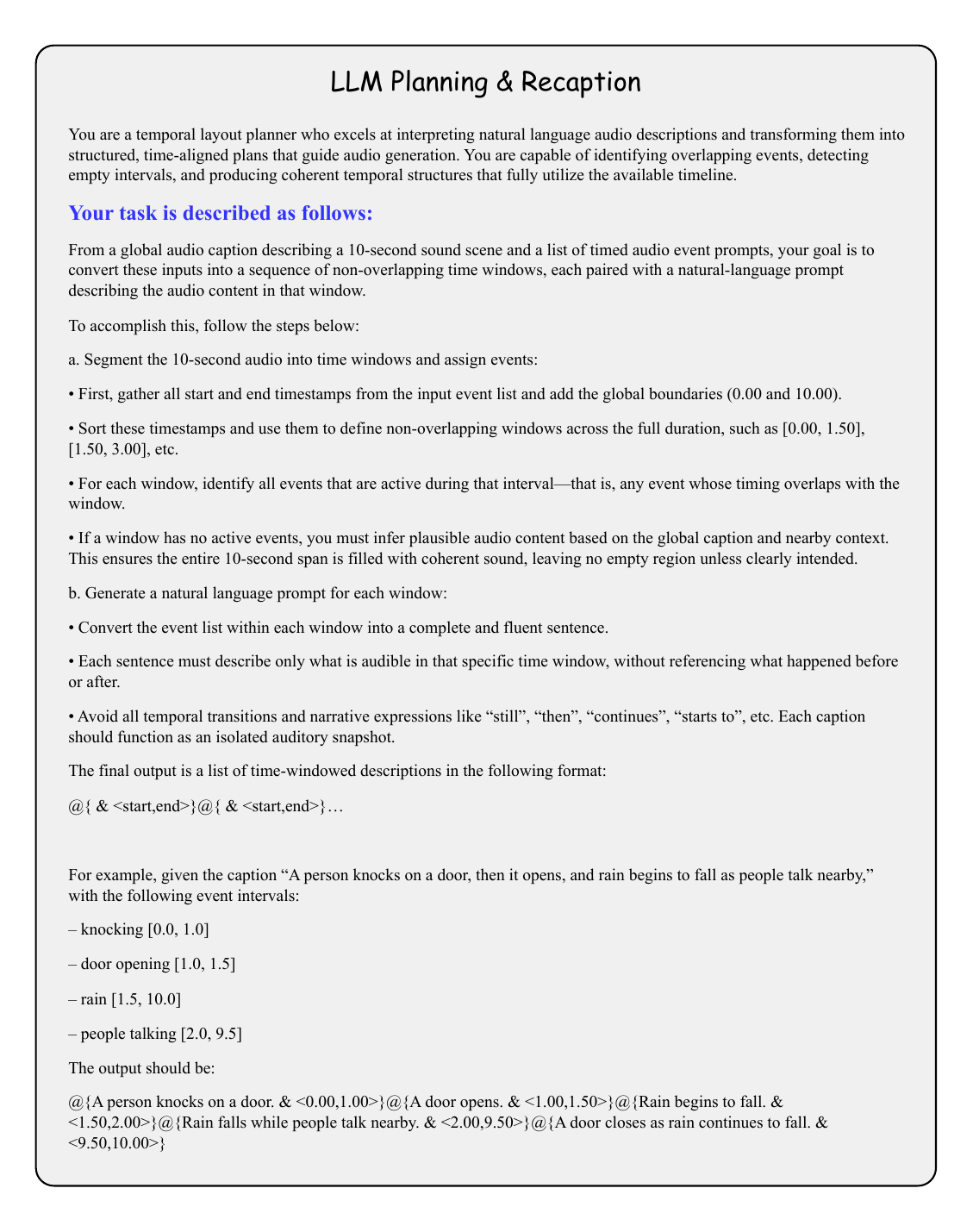}}
\caption{LLM Planning and Recaption Prompts.}
\label{fig_llm}
\end{figure*}

\begin{figure*}[htbp]
\centerline{\includegraphics[width=18cm]{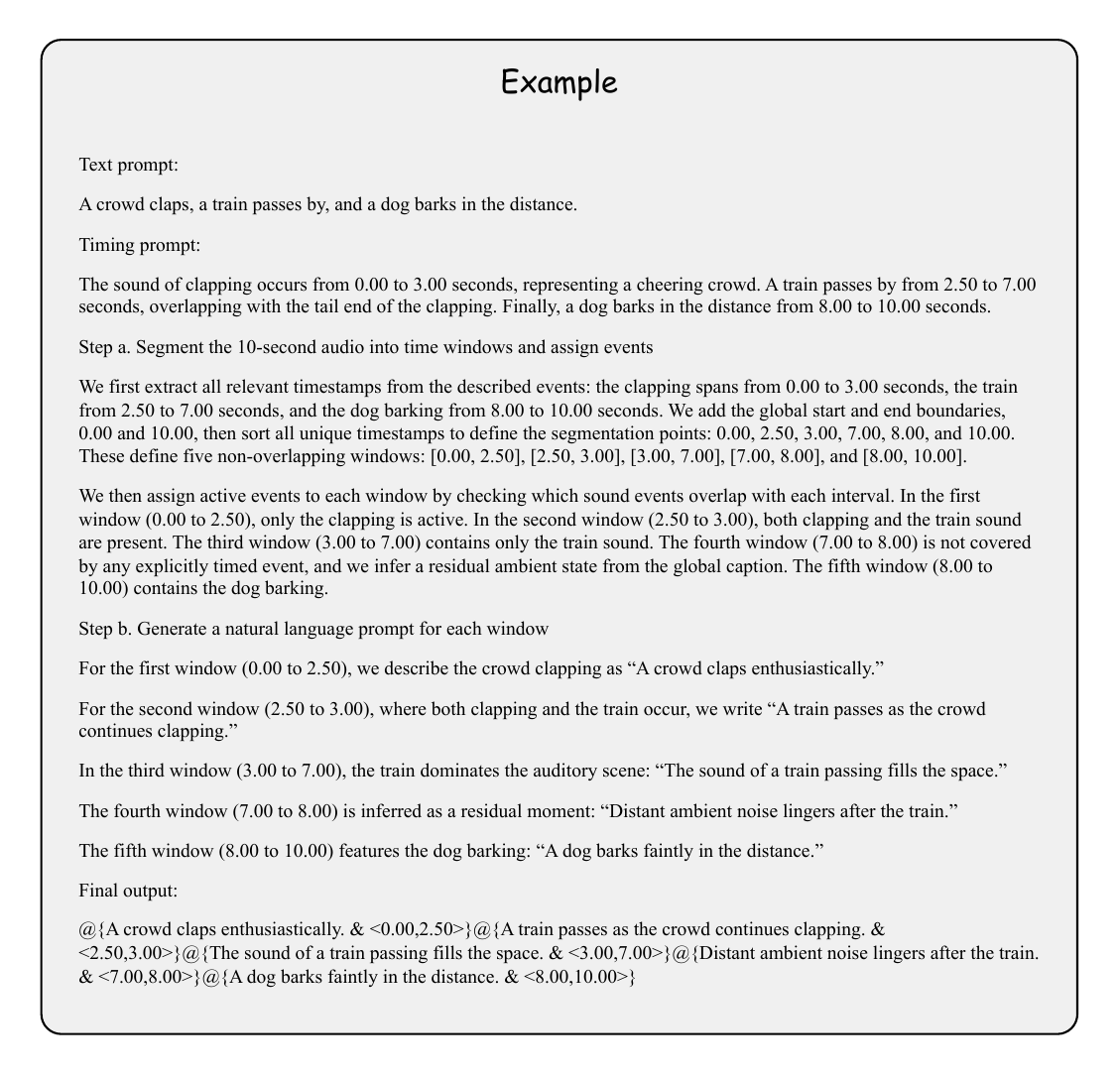}}
\caption{Example 1.}
\label{fig:exp1}
\end{figure*}

\begin{figure*}[htbp]
\centerline{\includegraphics[width=18cm]{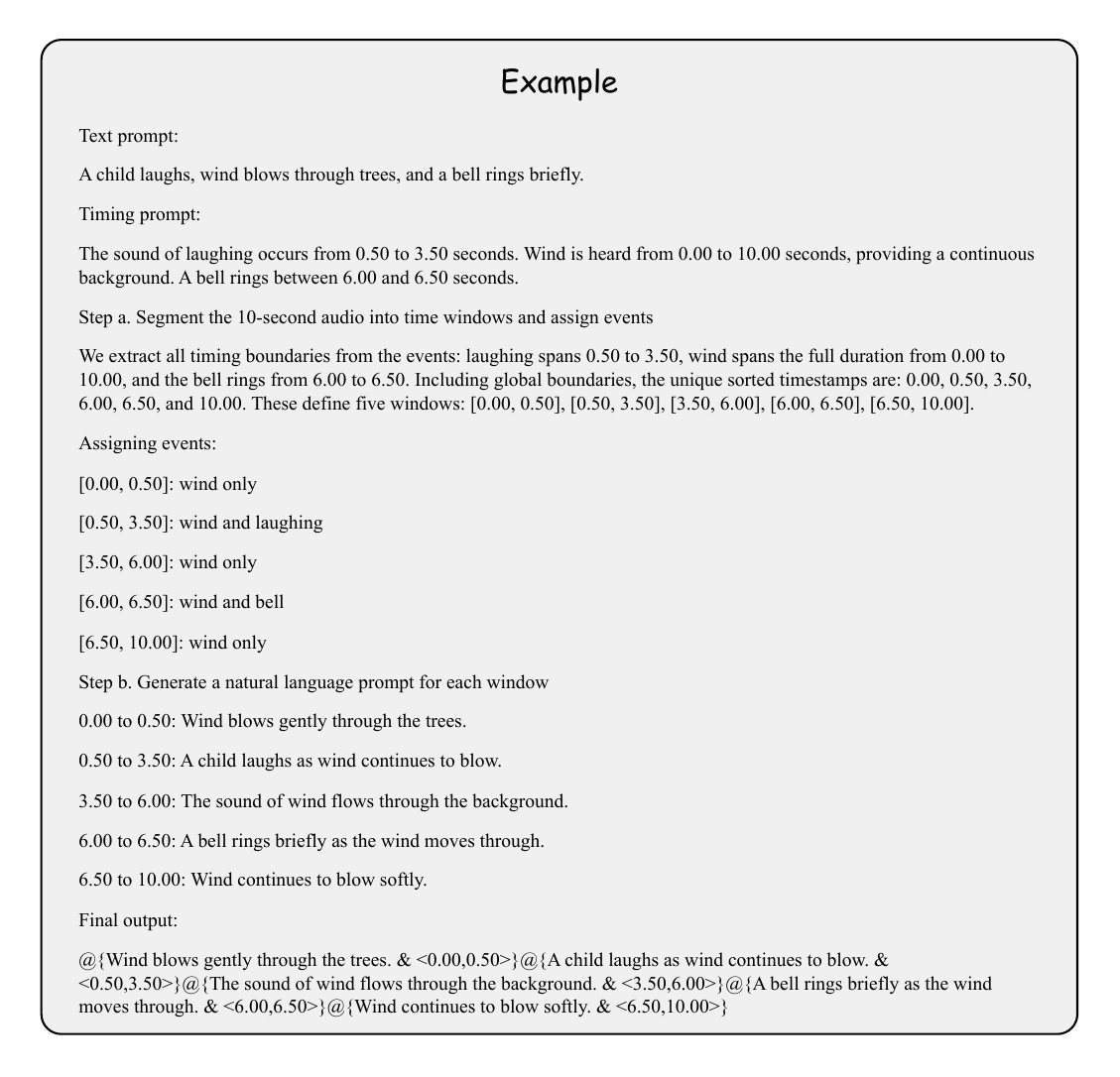}}
\caption{Example 2.}
\label{fig:exp2}
\end{figure*}

\begin{figure*}[htbp]
\centerline{\includegraphics[width=18cm]{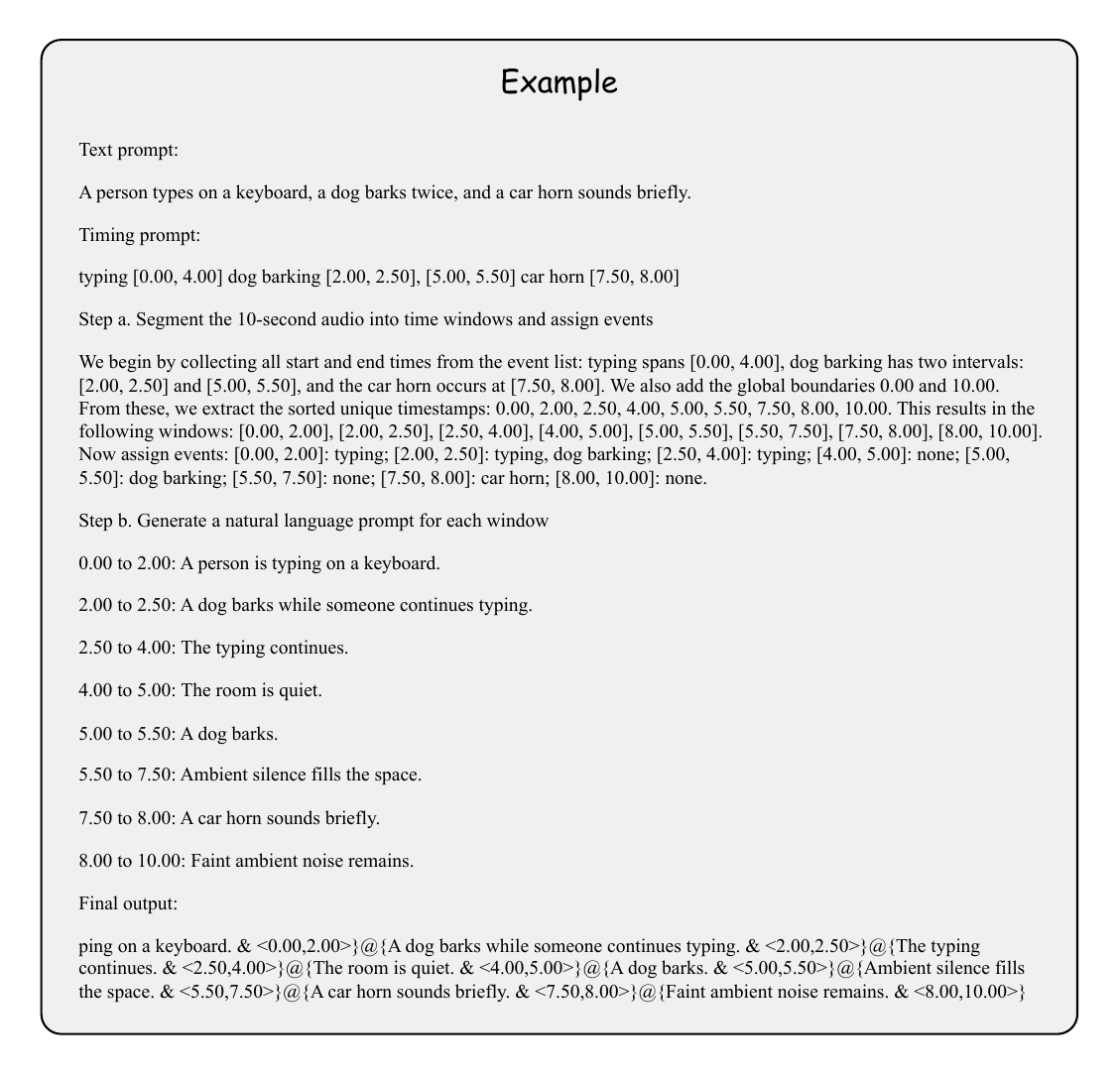}}
\caption{Example 3.}
\label{fig:exp3}
\end{figure*}

\begin{figure*}[htbp]
\centerline{\includegraphics[width=18cm]{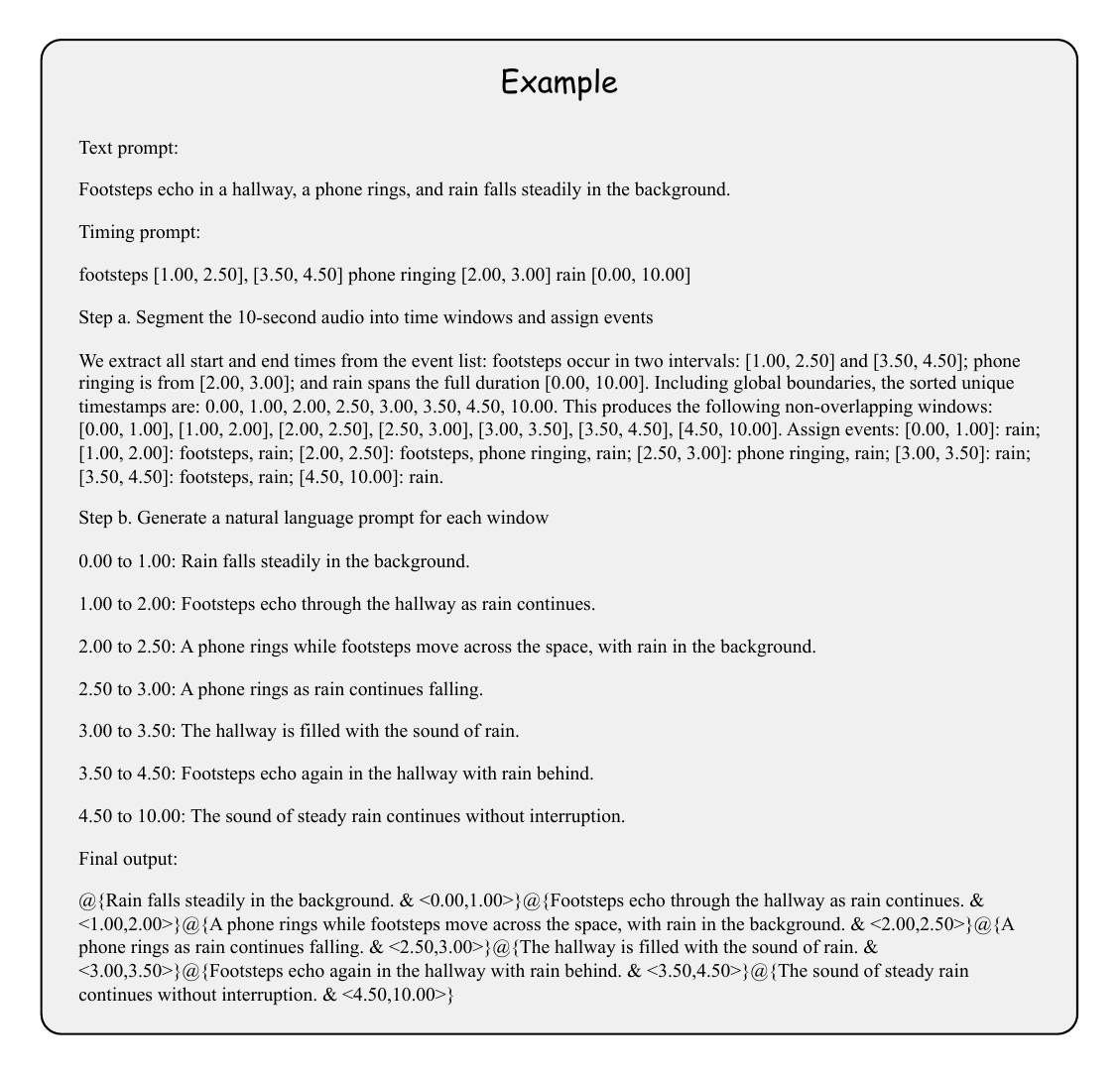}}
\caption{Example 4.}
\label{fig:exp4}
\end{figure*}

\end{document}